\documentclass[10pt,aps,prd,twocolumn, nofootinbib]{revtex4-1}

\usepackage{amssymb,amsmath,accents}
\usepackage{subcaption,verbatim}
\usepackage{graphicx}
\usepackage{color,units}
\usepackage{hyperref}
\usepackage{listings}
\usepackage[normalem]{ulem}
\hypersetup{
    bookmarks=true,         
    unicode=false,          
    pdftoolbar=true,        
    pdfmenubar=true,        
    pdffitwindow=false,     
    pdfstartview={FitH},    
    pdfauthor={Boris Goncharov},     
    colorlinks=true,       
    linkcolor=blue,          
    citecolor=red,        
}
\graphicspath{{figures/}}

\begin{document}

\definecolor{dkgreen}{rgb}{0,0.6,0}
\definecolor{gray}{rgb}{0.5,0.5,0.5}
\definecolor{mauve}{rgb}{0.58,0,0.82}

\lstset{frame=tb,
  	language=Matlab,
  	aboveskip=3mm,
  	belowskip=3mm,
  	showstringspaces=false,
  	columns=flexible,
  	basicstyle={\small\ttfamily},
  	numbers=none,
  	numberstyle=\tiny\color{gray},
 	keywordstyle=\color{blue},
	commentstyle=\color{dkgreen},
  	stringstyle=\color{mauve},
  	breaklines=true,
  	breakatwhitespace=true
  	tabsize=3
}

\title{All-sky radiometer for narrowband gravitational waves using folded data}
\author{Boris Goncharov}%
 \email{boris.goncharov@ligo.org}
\affiliation{Monash Centre for Astrophysics, School of Physics and Astronomy, Monash University, VIC 3800, Australia}
\affiliation{OzGrav: The ARC Centre of Excellence for Gravitational-Wave Discovery, Clayton, VIC 3800, Australia}
\author{Eric Thrane}
\affiliation{Monash Centre for Astrophysics, School of Physics and Astronomy, Monash University, VIC 3800, Australia}
\affiliation{OzGrav: The ARC Centre of Excellence for Gravitational-Wave Discovery, Clayton, VIC 3800, Australia}
\date{\today}

\begin{abstract}
We demonstrate an all-sky search for persistent, narrowband gravitational waves using mock data. The search employs radiometry to sidereal-folded data in order to uncover persistent sources of gravitational waves with minimal assumptions about the signal model. The method complements continuous-wave searches, which are finely tuned to search for gravitational waves from rotating neutron stars, while providing a means of detecting more exotic sources that might be missed by dedicated continuous-wave techniques.
We apply the algorithm to simulated Gaussian noise.
We project the strain amplitude sensitivity assuming circularly polarized signals for the LIGO network in the first observing run to be $h_0 \approx 1.2 \times 10^{-24}$ (1\% false alarm probability, 10\% false dismissal probability).
We include a treatment of instrumental lines and detector artifacts using time-shifted LIGO data from the first observing run.
\end{abstract}

\maketitle

\section{\label{sec:intro} Introduction}
With the first observations of a binary neutron star inspiral GW170817~\cite{GW170817firstBNS} and multiple black hole mergers~\cite{GW150914,GW151226,GW170104,GW170814,GW170608} by Advanced LIGO~\cite{2010advancedLIGO} and Advanced Virgo~\cite{2014advancedVirgo}, it is clear that nature provides us with a unique way to study electromagnetically invisible processes using gravitational radiation.
The discovery of persistent gravitational-wave emission remains an interesting prospect for gravitational-wave astronomy.
In this work, we develop a method for detecting quasi-monochromatic, persistent gravitational waves from unknown sources using data from advanced detectors.

Searches for continuous gravitational waves are designed to be as sensitive as possible to rotating neutron stars. However, to achieve this, they employ a highly tuned signal model.
If neutron stars emit gravitational waves in a way that does not match standard models, or if there are exotic sources of persistent gravitational waves, the signal could be missed by current continuous-wave searches.
The radiometer~\cite{ballmer2006radiometer,thrane2009anisotropicgwbg,radiometer2007s4,radiometer2011s5,stochastic2017directional} provides a solution.
By cross-correlating data from two or more detectors, it is possible to discover weak signals without a model for the signal phase evolution. Due to computational limitations, previous radiometer searches were either targeted (pointing in one direction) and narrowband (considering many different frequencies) or all-sky (looking in all directions) but broadband (averaging over all frequencies).
Since it seems unlikely that point sources of persistent gravitational waves would be broadband, it is desirable to carry out an all-sky narrowband search\footnote{After this paper was submitted for publication, a preprint appeared proposing gravitational waves from networks of primordial black holes connected by strings~\cite{Vilenkin}. It seems possible to us that such a network could produce broadband point sources.}. 
In~\cite{thrane2015folded} it was pointed out that sidereal folded data~\cite{ain2015folding} can be used to carry out a computationally cheap search that is both all-sky and narrowband.
In this paper, we employ the method from~\cite{thrane2015folded} to demonstrate the technique on an end-to-end study of Monte-Carlo noise.
Using limited data from LIGO's first observing run (O1), we show how vetoes can be used to manage instrumental artifacts found in real data.

The rest of the paper is organized as follows. Section~\ref{sec:motivation} provides the motivation for a search for unmodeled persistent sources.
In Section~\ref{sec:method}, we provide an overview of the narrowband radiometer with folded data. Section~\ref{sec:dataquality} describes how we handle instrumental artefacts and other data quality issues. In section~\ref{sec:detstat}, we demonstrate the detection of simulated signals. In Section~\ref{sec:sensitivity}, we calculate the sensitivity of the search. 

\section{\label{sec:motivation} Motivation}
\subsection{\label{sec:astromotivation} Astrophysical sources}
Accreting neutron stars in binaries are considered to be promising candidate sources of persistent gravitational waves. Optimistic models predict for such systems to have an asymmetrical quadrupole moment of inertia due to either deformation of the stellar interior~\cite{melatos2005deformInterior} or localized mass accumulation~\cite{vigelius2009localMassAccum}.
In either scenario, the quadrupole moment of inertia evolves through accretion and the influence of the neutron star magnetic field.
The quadrupole moment may be sustained even when accretion has abated.
Gravitational-wave driven instabilities of $r$-mode oscillations are another source of a quadrupole moment~\cite{reisenegger2003rmodes, ushomirsky2001rmodes}. If accretion is persistent, and neglecting torque from gravitational waves emission, neutron stars are expected to eventually spin up to their break up frequency $f \approx\unit[1400]{Hz}$~\cite{chakrabarty2003maxSpinFreq}.
However, the highest yet observed frequency is $f \approx 716$ Hz for a millisecond pulsar~\cite{hessels2006pulsObserv} and $f \approx 600$ Hz for an accreting millisecond pulsar~\cite{patruno2012accretingPulsObserv}, which is consistent with a hypothesis that emission of persistent gravitational waves prevents further spin up of neutron stars.
This is known as the torque balance hypothesis~\cite{wagoner1984nsgw, bildsten1998nsgw}.

Searches for continuous gravitational waves specifically target neutron stars.
However, in certain circumstances these searches can be sub-optimal.
For example, when a neutron star is in a binary system, it is computationally challenging to search the full signal parameter space.
Another example is a neutron star glitch, a sudden increase in the rotation frequency, a phenomenon observed in the timing of many radio pulsars~\cite{espinoza2011glitchescatalogue}. It has been shown that neutron star glitches can cause a loss of a substantial fraction of a signal-to-noise ratio in continuous wave searches~\cite{ashton2017cwglitches}.

Another motivation for the method discussed here is to explore the possibility of unknown persistent and narrowband signals.
One such theoretical scenario is gravitational waves from super-radiance of massive clouds of ultralight axions around a Kerr black hole~\cite{east2017BHSuperRadiance}. 
The frequency of gravitational waves from this long-lived resonance depends on the mass of a hypothetical axion particle. Thus, a narrowband emission is expected. Axions with a mass of $\sim \unit[10^{-11}-10^{-14}]{eV}$ could possibly be detected by Advanced LIGO~\cite{brito2017ultralightBosons}.

\subsection{\label{sec:othersearches} Searches for persistent gravitational waves}
Currently there are several methods for persistent gravitational wave searches. In this section of the paper, we outline what niche the narrowband radiometer search occupies.
A comprehensive overview of current searches for persistent gravitational waves can be found in~\cite{riles2017CWoverview}.

One of the main difficulties in searches for persistent gravitational wave is the amount of computational resources that are required to probe the parameter space of possible gravitational-wave frequencies and their time derivatives.
Searches have to account for Doppler modulations of the gravitational-wave signal due to motion of the Earth. Moreover, torque exerted on a neutron star by accretion from a companion star may change with time, resulting in wandering of the neutron star spin frequency~\cite{ghosh1979accretion}.

Knowing orbital parameters for some sources of persistent gravitational wave emission eliminates the problem of searching over gravitational-wave frequencies and their derivatives. Using the data from radio and gamma-ray observations, recent searches placed upper limits on gravitational-wave strain from 200 known pulsars~\cite{cw2017knownpulsars}. If the target is an accreting neutron star in a binary system, it may be possible to narrow down a parameter space by looking at X-ray pulsations~\cite{galloway2013ScoX1pulsations}.

There are three kinds of searches.
Searches for neutron stars with known sky position and known frequency are referred to as targeted searches.
Directed searches target specific sky locations without assumptions about the gravitational wave frequency.
All-sky searches employ no assumptions for either sky location or frequency.
Targeted searches can employ matched filtering~\cite{jaranowski1998matchedfilter}, a Bayesian approach~\cite{dupuis2005bayesian}, and the ``Five-vector'' method~\cite{astone2010fivevector}.

In a directional search one faces a problem of exploring a vast parameter space of  frequency and its derivatives.
Fully coherent searches are too computationally expensive for all-sky searches, and are adapted to limited observation time and/or specific sky directions ~\cite{wette2008fullycoherent, dhurandhar2008crosscorr}. In semi-coherent searches one instead sums results from coherent analysis over much shorter time intervals, for longer observation time~\cite{wette2015semicoh}. Semi-coherent methods are less computationally expensive than fully coherent ones, and sometimes they are used for all-sky searches. TwoSpect is an example of a template-based semi-coherent all-sky search, which tracks Earth's rotation-induced modulations of gravitational waves in doubly Fourier transformed data~\cite{goetz2011twospect}.
Polynomial algorithm uses a bank of frequency polynomials for matched filters~\cite{putten2010polynomial}. Hidden Markov model tracking method using a Viterbi algorithm~\cite{suvorova2016hmmViterbi} for matched filtering.
Other semi-coherent searches include ``Stack Slide''~\cite{brady1998stackslide}, the Hough approach~\cite{hough1959proceedings, krishnan2004skyhough, antonucci2008freqhough}, Powerflux~\cite{dergachev2010powerflux}, and Einstein@Home~\cite{abbott2009einsteinhome}, a volunteer-distributed computing project.
These semi-coherent search strategies rely on signal models of gravitational-wave emission from neutron stars.
They are in some sense limited by computational resources.

A different approach to the problem of a frequency modulated signal is to formulate a model-independent search. The radiometer technique is used to identify signals with the cross-correlation of Fourier-transformed strain from two or more gravitational wave detectors~\cite{ballmer2006radiometer}, and it underpins the method described in this paper. Cross-correlation contains information about the source sky location.
The radiometer works with minimum assumptions about a signal, only presuming it is persistent and narrowband.

Scorpius X-1 is the second brightest persistent X-ray source in the sky~\cite{giacconi1962scox1}. It is believed to be an accreting neutron star. According to the torque balance hypothesis, this system is a source of persistent gravitational waves. Using simulated Scorpius X-1 signals, it has been estimated that the {\em targeted} radiometer algorithm has less sensitivity than CrossCorr, a comparable sensitivity to TwoSpect, while at the same time it uses less than 1\% of computational resources of these pipelines~\cite{ScoX1Comparison15}. It has been demonstrated that a lossless data compression technique called folding can complement the narrowband radiometer, further reducing a computational cost and solving a data storage problem ~\cite{thrane2015folded}.
By combining folding with radiometry, we seek to extend the radiometer to carry out a computationally efficient all-sky search.

\section{\label{sec:method} Method}
In this section we describe a procedure of transforming a gravitational wave strain $s_{1,2}(t)$ measured by two interferometers into the radiometer signal-to-noise ratio, which will serve as the basis for our detection statistic.
In subsection~\ref{sec:crosscorrelation} we explain the process of cross-correlation.
Subsection~\ref{sec:datafolding} describes an implementation of data folding. In subsection~\ref{sec:radiometer} we apply the directional narrowband radiometer on a folded dataset.

\subsection{\label{sec:crosscorrelation}Cross-correlation}
Following the procedure from~\cite{thrane2015folded}, we divide the data into discrete segments indexed by start time $t$\footnote{
We use the variable $t$ to denote both segment start time and sampling time.
The meaning of any particular $t$ should be clear by context.}.
There are important considerations that determine a suitable choice of segment duration.
On one hand, longer time segments lead to a better frequency resolution.
On the other hand longer segment duration decreases search sensitivity at high frequencies due to the rotation of the Earth; see Eq.~12 and the surrounding discussion in~\cite{thrane2015longtransients}.
The overall range of frequencies we consider is between 20 Hz and 1800 Hz.
The maximum frequency imposes requirements on our segment duration.
In this analysis we pick a segment duration time for the Fourier transformation to be 32 s.
This choice guarantees  $<5\%$ decrease in the signal to noise ratio at $\unit[1800]{Hz}$ due to the rotation of the Earth.

For each segment, we calculate the Fourier transform of the strain time series $\tilde{s}_{1,2}(t,f)$.
The subscript refers to the detector number.
Noise power spectral densities for each individual detector $P_{1,2}$ are calculated for the background estimation using adjacent time segments. Next, complex-valued estimators $\upsilon(t,f)$ and $\sigma(t,f)$ are computed for each sidereal day of the observation:
\begin{equation}\label{eq:crosscorrsignal}
\upsilon(t,f) \equiv \frac{1}{N} \tilde{s_1}^{*}(t,f) \tilde{s_2}(t,f) \in \mathbb{C}
\end{equation}
\begin{equation}\label{eq:crosscorrsigma}
\sigma (t,f) \equiv \frac{1}{2} \sqrt{P_1^{'}(t,f) P_2^{'}(t,f)} \in \mathbb{R}
\end{equation}
\begin{equation}\label{eq:crosscorrsnr}
\rho(t,f) \equiv \frac{\upsilon(t,f)}{\sigma (t,f)} \in \mathbb{C}
\end{equation}
In Eq~\ref{eq:crosscorrsignal}, $N$ is a normalization constant defined in \cite{ballmer2006radiometer}.
It is introduced so that $\upsilon(t,f)$ has units of power spectral density.
Note that $\upsilon(t,f)$ is equivalent to $\mathfrak{Y}$ in~\cite{thrane2015folded}.

Next, we apply a coarse-graining operation~\cite{thrane2015longtransients} by combining neighboring frequency-domain points of $\upsilon(t,f)$ and $P_{1,2}^{'}$:
\begin{equation}\label{eq:coarsegrain1}
\upsilon(t,f_\text{CG}) = \frac{1}{q}\sum\limits_{i = p}^{p+q-1} \upsilon(t,f_{i})
\end{equation}
\begin{equation}\label{eq:coarsegrain2}
\\P^{'}_{1,2}(t,f_\text{CG}) = \frac{1}{q}\sum\limits_{i = p}^{p+q-1} P^{'}_{1,2}(t,f_{i}) ,
\end{equation}
where
\begin{align}
f_\text{CG}=\frac{1}{2}\left(f_i + f_{i+q-1}\right) .
\end{align}
Choosing the degree of coarse-graining is a balancing act like choosing the segment duration.
If we make the coarse-grained bins too wide, we needlessly add noise on top of the signal.
If the coarse-grained bins are too small, the signal may wander outside of the bin.
In this analysis, we coarse-grain power spectra from an intrinsic frequency resolution of $1/32$ Hz to 1 Hz.
Varying the coarse-grained resolution to consider a variety of scenarios is possible, but this lies outside our present scope. 

\subsection{\label{sec:datafolding} Data folding}
Due to the rotation of the Earth, the expectation value of $\rho(t | f)$ for a persistent narrowband signal at a frequency $f$ is a periodic function, with a period equal to one sidereal day.
Folding is a data compression technique that uses this symmetry to transform any dataset into only one sidereal day of data~\cite{ain2015folding}.

\begin{figure}[h]
\centering
\includegraphics[width=1.0\linewidth]{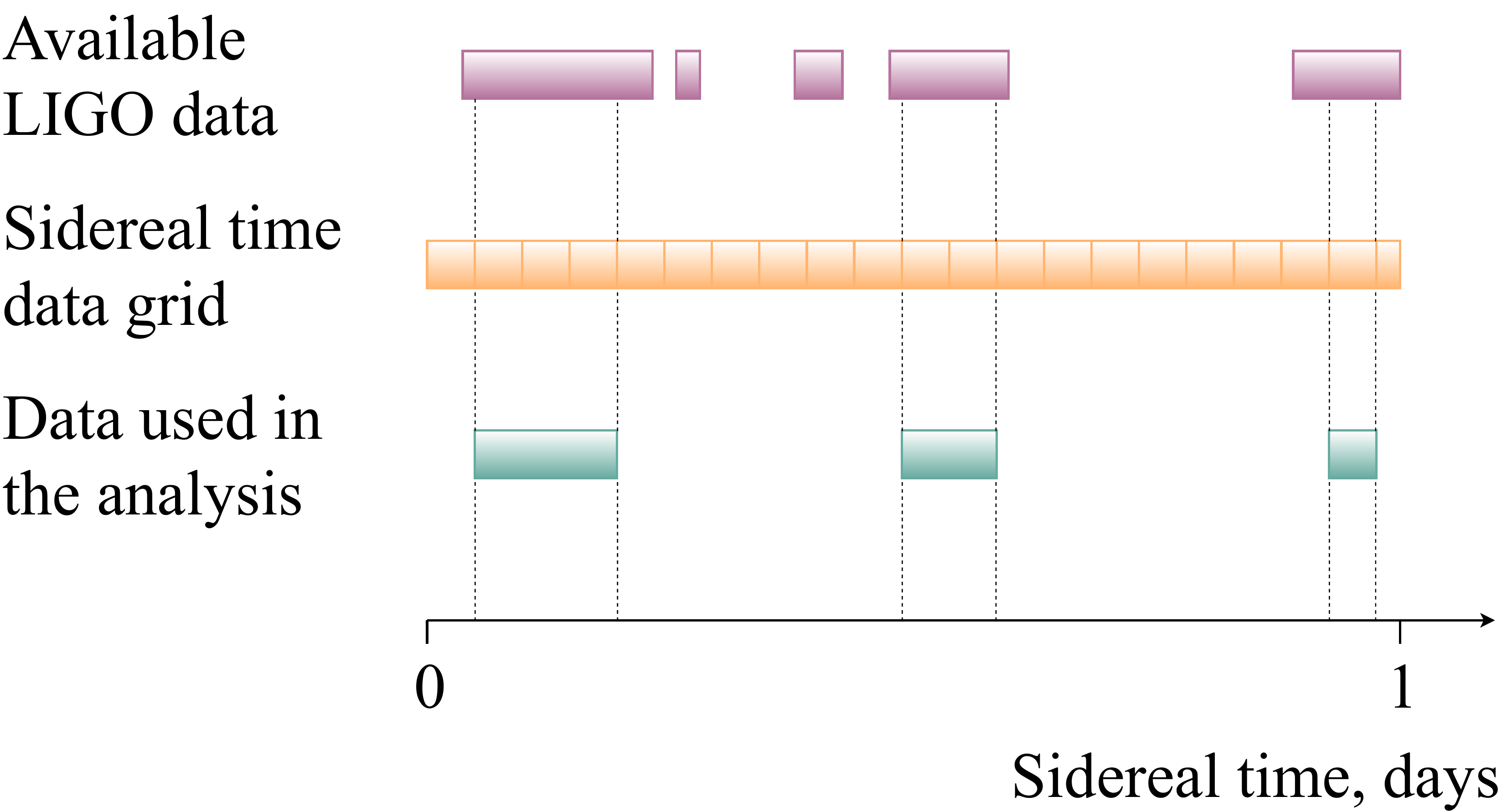}
\caption{An illustration of how LIGO data is arranged into segments prior to the folding operation.}
\label{fig:sidereal-grid}
\end{figure}

First, we select a GPS time that  corresponds to sidereal time = 0 for the first sidereal day.
We define an array of evenly-spaced, $\unit[32]{s}$ segments starting from this zero time.
Segments that overlap between sidereal days are removed.
The first complete segment in a sidereal day becomes the first segment of that day.

Fragments of data that do not fit into the new time segments are truncated (Figure \ref{fig:sidereal-grid}). Interferometer lock segments shorter than 700 seconds are removed as well.
Applying these cuts to data from LIGO's first observing run, approximately $6\%$ of the data is removed.

Next, following~\cite{thrane2015folded}, we sum over sidereal days $k$ in order to fold the data into just one sidereal day $\rho_{\text{fold}}(f,t | k)$, using $\sigma(f,t)$ as weight coefficients:

\begin{equation}\label{eq:yFolding}
\upsilon_{\text{fold}}(t,f | k) = \frac{\sum_{k} \upsilon_k(t,f) \sigma_k^{-2}(t,f) }{\sum_{k} \sigma_k^{-2}(t,f) }
\end{equation}

\begin{equation}\label{eq:sigmaFolding}
\sigma_{\text{fold}}(t,f | k) = (\sum_{k} \sigma_k^{-2}(t,f))^{-\frac{1}{2}}
\end{equation}

\begin{equation}\label{eq:pFolding}
\rho_{\text{fold}}(t,f | k) \equiv \frac{\upsilon_{\text{fold}}(t,f)}{\sigma_{\text{fold}}(t,f)}
\end{equation}

\begin{figure}[h]
\centering
\includegraphics[width=1.0\linewidth]{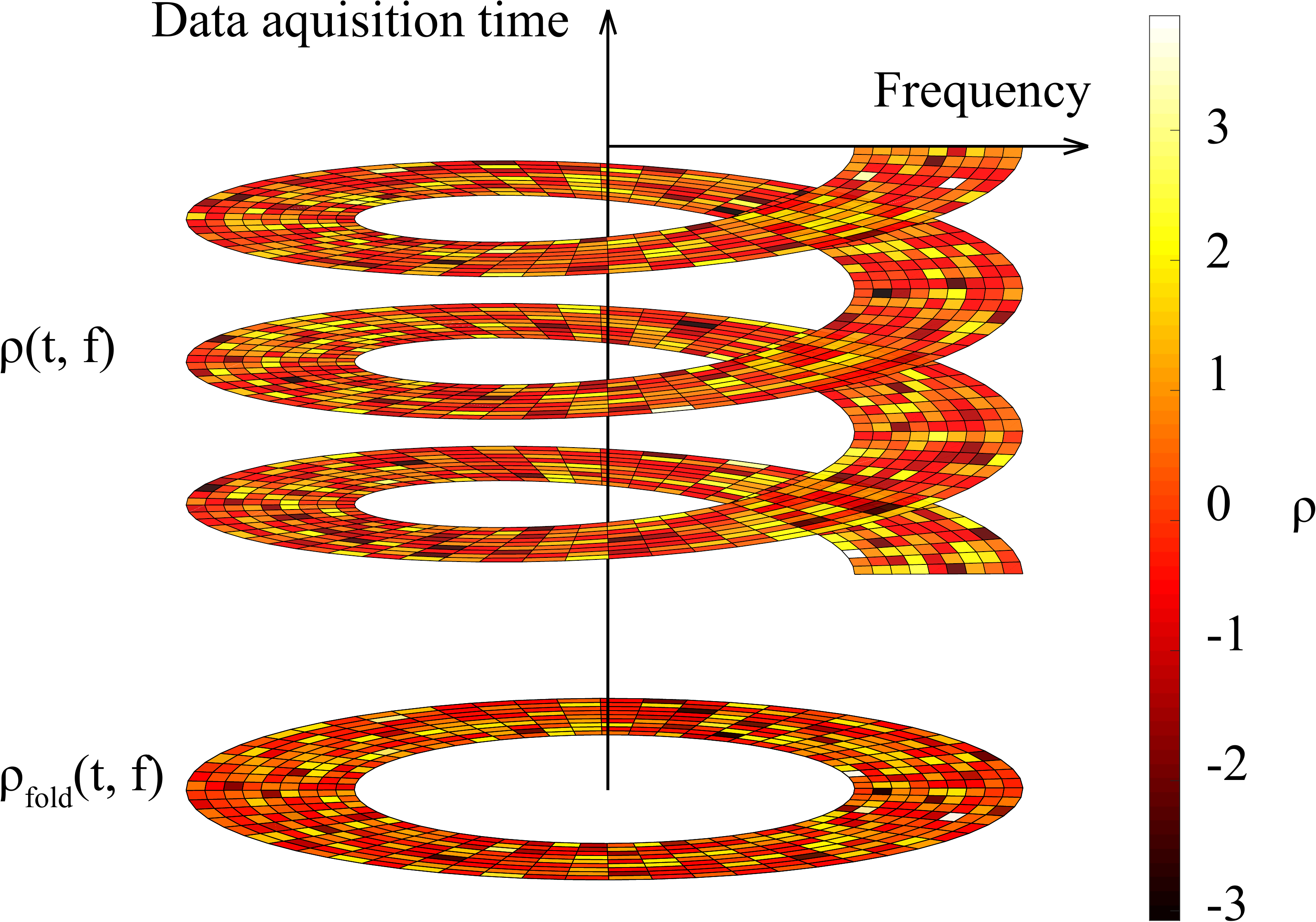}
\caption{Representation of data folding. Each element of the helix with a fixed radius represents a real part of the $\rho_{\text{fold}}(t | f)$ data set at a fixed frequency, while each revolution of the helix represents one sidereal day of observations. The ring below the helix represents a folded dataset $\rho_{\text{fold}}(t,f | k)$, where each element is calculated on a basis of the above cells of the helix.}
\label{fig:foldingvisual}
\end{figure}

\subsection{\label{sec:radiometer}Radiometry}
Gravitational-wave radiometry~\cite{ballmer2006radiometer} has been used in searches for persistent gravitational waves~\cite{abbott2007firstradiometer,radiometer2011s5,stochastic2017directional}. The first LIGO radiometer analysis was carried out in 2007~\cite{abbott2007firstradiometer}. Narrowband radiometry provides us with a spectrum of a gravitational wave strain data at each sky location. Following~\cite{ballmer2006radiometer,thrane2015folded}, the signal-to-noise ratio is given by:
\begin{equation}\label{eq:radiometer}
\text{SNR}(f | \hat{\mathbf{\Omega}}) = \frac{ \sum_{t} \text{Re}(\rho_{\text{fold}}(t; f) e^{2 \pi i f \hat{\mathbf{\Omega}} \cdot \Delta \vec{\mathbf{x}}(t)/c} ) \epsilon_{12} (t | \hat{\mathbf{\Omega}}) }{ \sqrt{\sum_{t} \epsilon_{12}^2 (t | \hat{\mathbf{\Omega}})} }
\end{equation}
Here $\hat{\mathbf{\Omega}}$ is the unit vector pointing to the sky position of the source, $\Delta \vec{x}(t)$ is the separation vector of detectors, c is the speed of light, and $\epsilon_{12} (t | \hat{\mathbf{\Omega}})$ is the sidereal-time-dependent efficiency factor.
\begin{equation}\label{eq:efficiency}
\epsilon_{12}(t | \hat{\mathbf{\Omega}}) \equiv \frac{1}{2} \sum_{A} F_1^A(t | \hat{\mathbf{\Omega}}) F_2^A(t | \hat{\mathbf{\Omega}})
\end{equation}
\\
\\Here $F_{1,2}^A(t | \hat{\mathbf{\Omega}})$ are antennae factors~\cite{hawking1989threeyearsgrav} for two interferometers; $A = [+,\times]$ are polarization states.

Previous studies using Monte-Carlo  data showed that radiometry with 20 days of folded data can be used to recover persistent gravitational wave signal with a strain amplitude $h_0 = 1.5 \times 10^{-24}$ at 600 Hz at the LIGO design sensitivity using two detectors with $\text{SNR} \approx 50$~\cite{thrane2015folded}.

\subsection{\label{sec:simsignal}Simulated signals}
To test the sensitivity of the algorithm, we simulate persistent gravitational waves.
Our simulated signals are circularly polarized with a fixed strain amplitude and a sinusoidally evolving phase.
The amplitude of the strain measured in each detector is modulated by the antenna factors, which change over the course of the sidereal day due to the rotation of the Earth; see, e.g.,~\cite{thrane2015folded}.
Injections are performed at an arbitrary fixed sky position $(\text{ra},\text{dec})=(\unit[21]{hr},9^\circ)$.
According to Figure 7b from~\cite{thrane2009anisotropicgwbg}, the radiometer sensitivity to strain power, averaged over a sidereal day, varies by about 40\% depending on the sky location.
Therefore we expect strain amplitude sensitivity to vary by about 20\% for different sky locations.
The simulated signals are injected into Gaussian noise corresponding to Advanced LIGO at design sensitivity~\cite{2010advancedLIGO}.
Technically, in order to simulate a signal with the characteristics of a continuous wave source in a binary, one ought to include time-dependent Doppler modulation. However, we ignore this effect in our simulation since our frequency bins are typically much wider than the expected Doppler modulation from binary motion.

\section{\label{sec:dataquality} Data quality}
Advanced LIGO comprises two detectors at Hanford and Livingston in the USA, and its first observing run (O1) took place between September 12, 2015 and January 19, 2016. 
It is necessary to remove instrumental lines to avoid false positives.
We provide a three-step technique to remove noise artifacts without accidentally removing an astrophysical signal.

The first step is to remove known instrumental lines from the frequency domain.
We employ a list of lines from the recent directed search for persistent gravitational waves using radiometry~\cite{stochastic2017directional}.

The second step is to remove times associated with non-stationary noise (glitches).
Since we are looking for a weak, persistent signal, we employ a relatively robust time-domain cut without fear of throwing out the signal.
Our time-domain cut eliminates any times that contain $N_\rho=6$ or more $\rho(t, f)-$spectrogram pixels with $\rho>(\rho_\text{max}=7)$.
This cut removes on average 0.5\% of O1 data (Figure ~\ref{fig:daycutstats}) and none of Monte-Carlo data.

While the first cut eliminates known lines (instrumental artifacts with known origins), there are additional ``unknown'' lines that we remove because they do not match our signal model.
The next step is to remove these unknown lines.
We apply an additional cut that eliminates any $\rho(t, f)-$spectrogram pixels with $\rho>\rho_\text{max}$.
This cut removes on average 0.3\% of the remaining O1 data and 0.1\% of the remaining Monte-Carlo data.
The values of $N_\rho$ and $\rho_\text{max}$ are chosen to produce real-data distributions of $\rho(t,f)$ that are comparable to distributions generated from Gaussian noise.

\begin{figure}[h]
\centering
\includegraphics[width=1.0\linewidth]{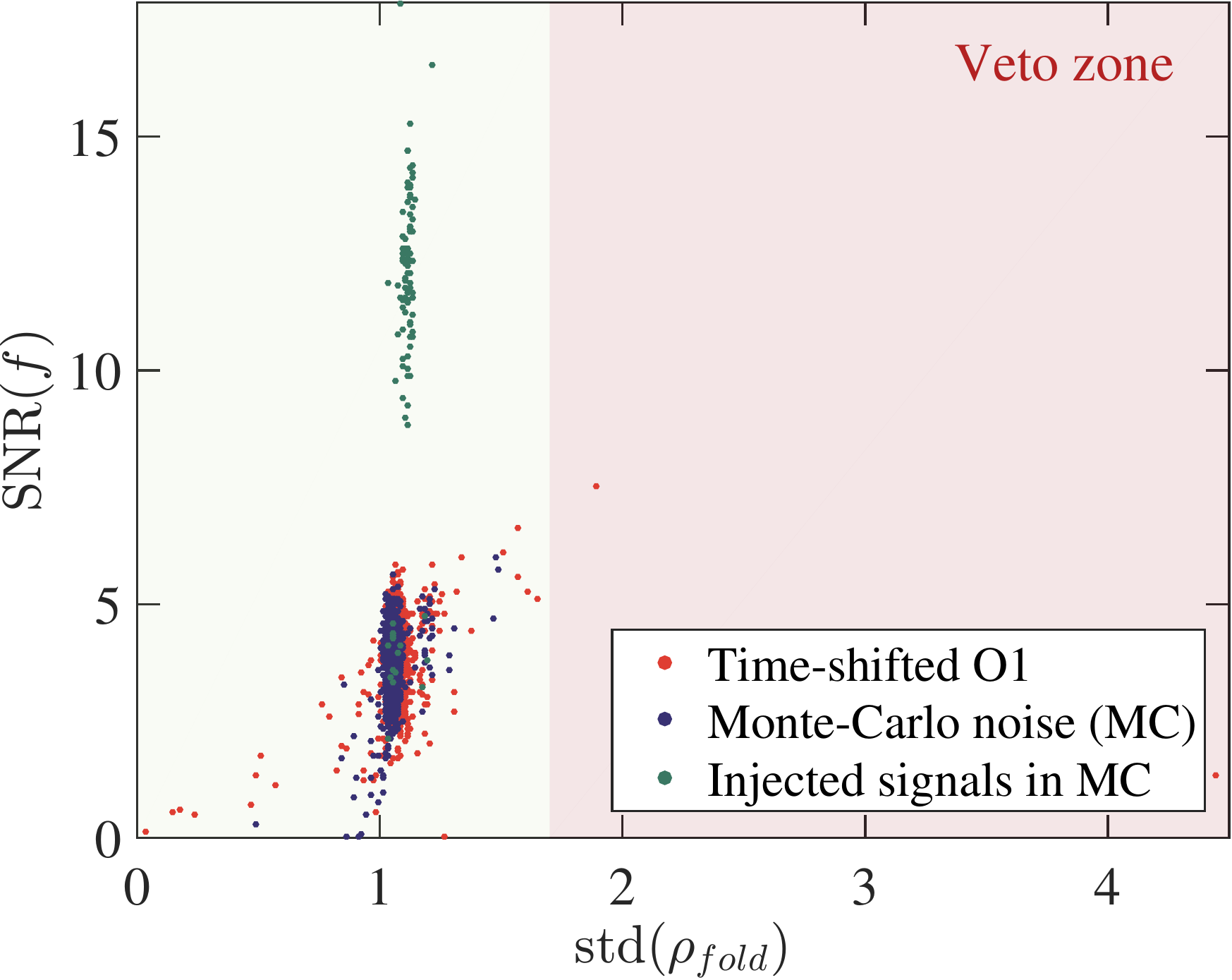}
\caption{The standard deviation of $\rho_{\text{fold}}(t | f,k)$ on the x-axis is used to veto frequency-domain data. The signal-to-noise ratio $\text{SNR}(f)$ on the y-axis quantifies significance. Some frequency bins in the time-shifted O1 data (red) with a high with a high $\rho_{\text{fold}}(t | f,k)$ in the veto zone of the plot would provide a great $\text{SNR}(f | \hat{\mathbf{\Omega}})$ if the veto were not applied.}
\label{fig:veto}
\end{figure}

Next, we look at the standard deviation of $\rho_{\text{fold}}(t | f,k)$ with respect to sidereal time, 
\begin{align}\label{eq:std}
\text{std}_t[\rho_{\text{fold}}(t | f,k)] .
\end{align}
While high values of $\rho_\text{fold}$ can be evidence of a signal, large {\em scatter} in the values of $\rho(t,f)$ is more likely to be due to a detector artifact.
We calculate the standard deviation in for each frequency bin as per Eq.~\ref{eq:std}.
Using injection studies, we set a maximum threshold on the standard deviation, which we denote $\sigma_\text{crit}$.
We determine that $\sigma_\text{crit} = 1.7$ is a suitable choice for vetoing instrumental artifacts while preserving signals.
This is illustrated in Fig.~\ref{fig:veto} and Fig.~\ref{fig:stdrerho}.
We plot the signal-to-noise ratio, maximized over all sky directions
\begin{equation}\label{eq:maxsnr}
\text{SNR}(f) = \max_{\hat{\mathbf{\Omega}}}\text{SNR}(f | \hat{\mathbf{\Omega}}) ,
\end{equation}
versus the standard deviation of $\rho_\text{fold}$ defined in Eq.~\ref{eq:std}.
Dots represent data from different frequency bins. 
Red represents time-shifted O1 data, blue represents Monte-Carlo noise, and green represents injected signals in Monte-Carlo noise. 
If the data falls into the red zone on the plot, it gets vetoed.

In addition to the above veto test using circularly-polarized signals, we perform an additional veto test using a linearly-polarized signal with an inclination angle of $\iota=90^{\circ}$ and a polarization angle of $\psi=0^{\circ}$.
By varying the integration time, we recover the signal multiple times with an $\text{SNR}$ between 4 and 50 at the injection frequency and sky location.
The veto threshold $\sigma_\text{crit}$ for the linearly-polarized signal is not exceeded.


\section{\label{sec:detstat} Detection statistic}
The goal of this section is to design a statistic for identifying the brightest point source on the sky and determining the associated statistical significance.
\\First, we find the brightest patch in the sky for all frequency bins, which we denoted $\text{SNR}(f)$. In this work we probe 360 equally-spaced radial components of angle $\hat{\mathbf{\Omega}}$ times 180 equally-spaced polar components for a a total of 64800 sky locations.

Next, we look for the frequency bin with the most significant $\text{SNR}(f)$.
Naively, one might expect that this is accomplished by choosing the maximum of $\text{SNR}(f)$ over all frequencies.
However, this naive method for finding the loudest frequency bins presumes that the noise distribution of $\text{SNR}(f)$ is independent of frequency.
In reality, the distribution changes as a function of frequency due to the fact that the diffraction limited resolution $\delta_\theta$ is a function of frequency:
\begin{equation}\label{eq:skyresolution}
\delta_\theta \approx \frac{c}{f} \frac{1}{\Delta \vec{x}(t)} \approx \bigg(\frac{1000 \ \text{Hz}}{f}\bigg) 5^\circ
\end{equation}
At high frequencies, there is a relatively higher number of effective sky locations.
Since there are more effective sky locations, fluctuations in the noise lead to greater $\text{SNR}(f)$ due to a trial factor effect.
We therefore must define a new statistic in order to avoid a  preference for higher frequency signals.



Our solution is to define a new statistic $\lambda(f)$, which rescales $\text{SNR}(f)$ to take into account the frequency dependence of the diffraction limited resolution:
\begin{equation}\label{eq:lambda}
\lambda(f) \equiv \frac{ \max_{\hat{\mathbf{\Omega}}}\text{SNR}(f,\hat{\mathbf{\Omega}}) - \mu_\text{fit}(f) }{\sigma_\text{fit}(f)} .
\end{equation}
The functions $\sigma_\text{fit}(f)$ and $\mu_\text{fit}(f)$ are measured empirically with simulations so that $\lambda(f)$ is approximately flat in frequency when we analyze noise.
The final detection statistic is
\begin{align}
\lambda \equiv \max_f \lambda(f) .
\end{align}

To assign a statistical significance to $\lambda$, we perform background simulations to generate a distribution of $\{\lambda_i\}$.
For each realization, we simulate an array of folded data $\rho_\text{fold}^i(t,f)$.
Every $(t,f)$ pixel is drawn from a normal distribution with mean=zero and with a variance determined from time-shifted data.
While individual segments of data are known to exhibit non-Gaussian noise, we expect that folded data to be nearly Gaussian distributed due to the central limit theorem.
This assumption is supported by previous cross-correlation analyses, e.g.,~\cite{radiometer2004s1,isotropic2007s4,radiometer2007s4,isotropic2012s5,radiometer2011s5,isotropic2014s6,stochastic2017isotropic,stochastic2017directional}.
We carry out $N_\text{sim}=10^5$ background realizations.
The false alarm probability (FAP) of $\lambda$ is given by:
\begin{equation}\label{eq:falsealarm2}
\text{FAP}(\lambda) = \frac{N(\lambda_i \geq \lambda)}{N_\text{sim}} ,
\end{equation}
where $N(\lambda_i\geq\lambda)$ is the number of simulated backgrounds that exceed $\lambda$ (the measured detection statistic).
In the remainder of the paper, we set $\text{FAP}=1\%$ as a fiducial threshold for identification of a statistically interesting signal.
We find that $\text{FAP}=1\%$ corresponds to a lambda value of $\lambda_0=7.6$.

\section{\label{sec:sensitivity} Sensitivity calculation}
In this section, we estimate the gravitational-wave strain amplitude $h_0(f)$ that we can detect with false alarm probability FAP = 1\% and false dismissal probability FDP = 10\%.
Our simulated signals are described in Section~\ref{sec:simsignal}.
For each frequency bin, we vary $h_0$ and determine the value such that we exceed the $\text{FAP}=1\%$ threshold $\lambda_0$ at least $1-\text{FDP}=90\%$ of the time.


\begin{figure}[h]
\centering
\includegraphics[width=1.0\linewidth]{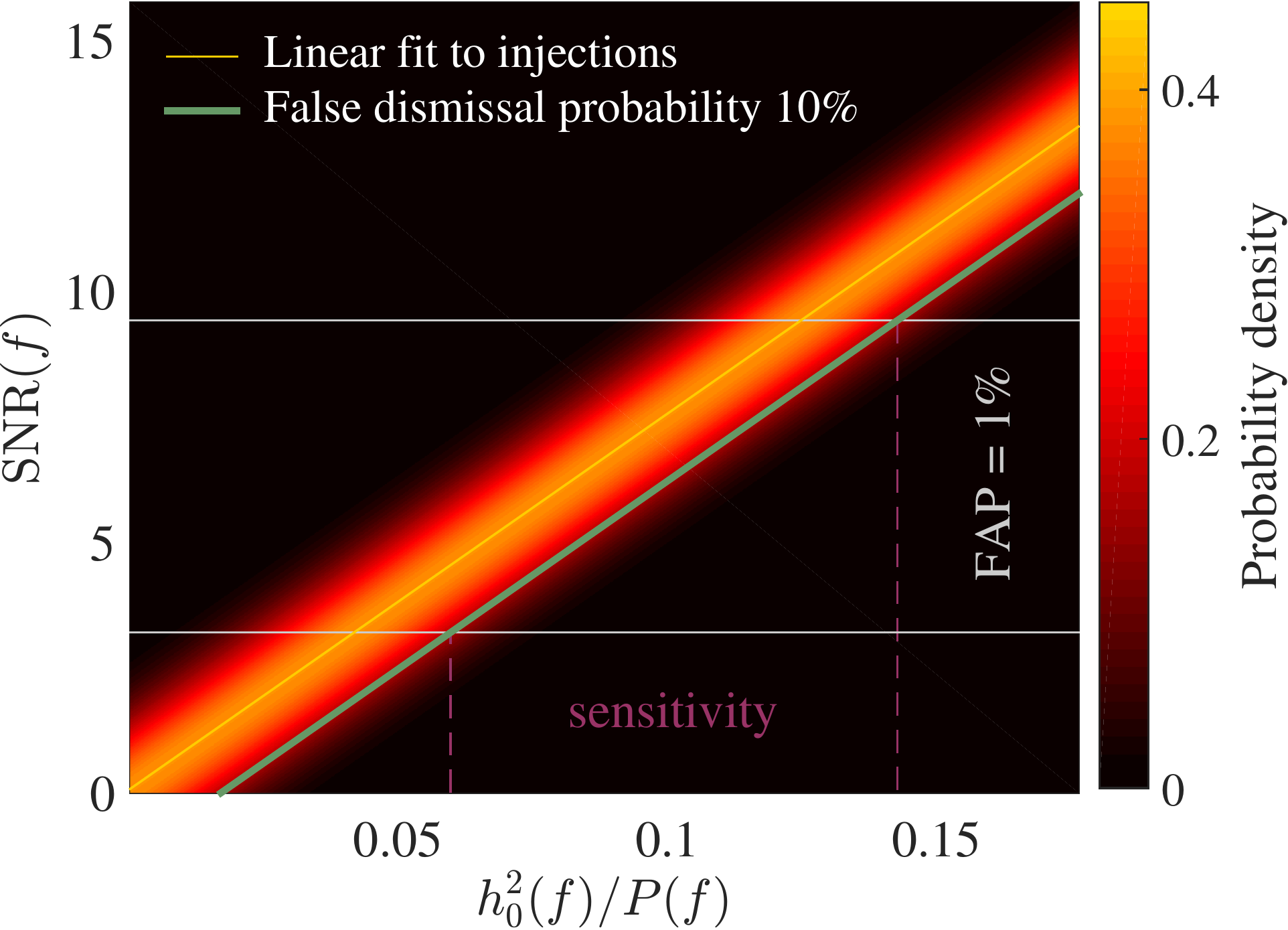}
\caption{Signal-to-noise ratio at the sky location where the signal was injected as a function of the effective injected signal to noise ratio. 
}
\label{fig:ulnew}
\end{figure}

In Figure~\ref{fig:ulnew}, we plot the recovered signal-to-noise ratio $\text{SNR}(f|\hat{\mathbf{\Omega}}_0)$ as a function of the effective injected signal-to-noise ratio $h_0^2(f) / P(f)$. 
%
The recovered signal-to-noise ratio is linearly proportional to the injected signal-to-noise ratio:
\begin{equation}\label{eq:snrscale}
\text{SNR}_m(f) = a \frac{h_0^2(f)}{P(f)} , 
\end{equation}
where $a$ depends on details of the windowing procedure, but for our choice of parameters, $a=74.9$.
For a fixed {\em injected} signal-to-noise ratio, there is a distribution of {\em recovered} signal-to-noise ratios, the width of which is indicated in Fig.~\ref{fig:ulnew} by the yellow-orange band.
The requirement that the false dismissal probability is $\text{FDP}=10\%$ can be visualized using the green line, below which 10\% of the injections are recovered for a fixed value of $h_0^2/P$.
We define $\Delta$ as the vertical distance between the yellow and green lines; it is the difference in SNR required to go from $\text{FDP}=50\%$ to $\text{FDP}=10\%$.
%
The strain amplitude sensitivity is
\begin{equation}\label{eq:ulh}
h_0 = \sqrt{\frac{P(f)}{a}\Big(\text{SNR}_0(f) - \Delta\Big)}  , 
\end{equation}
where $\text{SNR}_0(f)$ is the threshold for a statistically significant signal-to-noise ratio given $\lambda_0$.
That is,
\begin{align}
\text{SNR}_0(f) = \sigma_\text{fit}(f) \lambda_0 + \mu_\text{fit} .
\end{align}

\begin{figure}[h]
\centering
\includegraphics[width=1.0\linewidth]{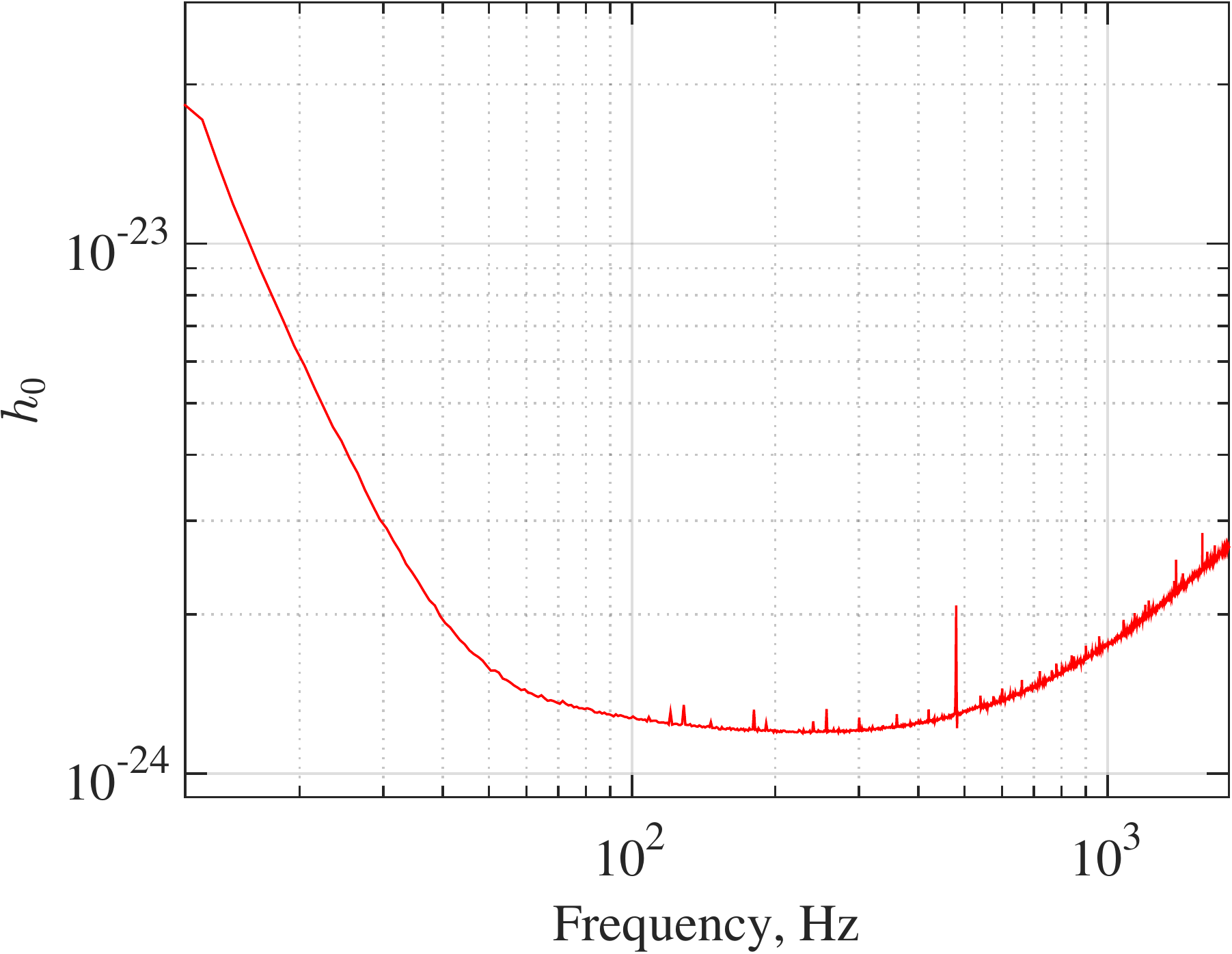}
\caption{Sensitivity to the strain amplitude $h_0(f)$ for 2 folded days of the Monte-Carlo background $\rho_{\text{fold}}(t,f | k)$ for LIGO at design sensitivity. We assume 1\% FAP and 10\% FDP.}
\label{fig:ulh}
\end{figure}

Our results are summarized in Fig.~\ref{fig:ulh}.
At the most sensitive frequency bin, corresponding to $f = 245 \text{Hz}$, the strain amplitude sensitivity is $h_0 = 1.3 \times 10^{-24}$.
For the current analysis with $\approx$ 2 sidereal days of Monte-Carlo noise at the level of LIGO design sensitivity we used a total of 4650 time segments 32 seconds long.
The signal-to-noise ratio scales like $\text{SNR}(f | \hat{\mathbf{\Omega}}) \propto \sqrt{t_{\text{obs}}}$~\cite{thrane2015folded}.
At $f = 600 \text{Hz}$ we project the strain amplitude sensitivity $h_0 \approx 3.9 \times 10^{-25}$, for a two-detector network operating at the level of LIGO design sensitivity for one year.
This prediction is the same order of magnitude as in~\cite{thrane2015folded} ($h_0 \approx 2 \times 10^{-25}$).
For analysis of LIGO's O1 run we project a strain amplitude sensitivity $h_0 \approx 1.2 \times 10^{-24}$ at $f = 245 \text{Hz}$. The expected sensitivity for linearly-polarized signals is expected to be worse by a factor of approximately 2.6~\cite{messenger2010radiometerpolarization}.

We compare this result to upper limits from recent searches for continuous and persistent gravitational radiation with LIGO's O1 data.
All-sky searches for continuous, nearly-monochromatic circularly-polarized gravitational waves in the 20-475 Hz band reported 95\% confidence upper limits that reach $h_0 \approx 1.5 \times 10^{-25}$ in the 150 - 250 Hz region~\cite{cw2017o1}.
Directional radiometer search using LIGO's O1 data provides 90\% confidence upper limits on persistent gravitational waves, reaching $h_0 \approx 4.0 \times 10^{-25}$~\cite{stochastic2017directional}.

\section{\label{sec:conclusion} Conclusion}
We apply an all-sky radiometer algorithm to simulated Gaussian noise, which has been compressed using sidereal-day folding.
The data are cleaned using a data-quality procedure developed with time-shifted data from LIGO's first observing run.
We project that the algorithm achieves a strain amplitude sensitivity of $\approx 1.2 \times 10^{-24}$ (1\% false alarm probability, 10\% false dismissal probability) for a two-detector network operating at design sensitivity for the time of the LIGO first observing run O1.
This corresponds to a sensitivity to neutron star ellipticity of \cite{74pulsars2007ligo}
\begin{equation}\label{eq:ellipticity}
\epsilon \approx 6 \times 10^{-5} \bigg(\frac{0.4}{\beta}\bigg) \bigg(\frac{10^{45}\ \text{g cm}^2}{I}\bigg) \bigg(\frac{r}{10\ \text{kpc}}\bigg) \bigg(\frac{600\ \text{Hz}}{f}\bigg)^2 , 
\end{equation}
where $\beta$ is an orientation factor, G is the gravitational
constant, r is the distance to the source, and I is the moment
of inertia.


Several improvements for narrowband searches with folded data can be implemented in the future.
Varying the coarse-grained frequency bin width can be used to achieve an optimal sensitivity for simulated signals.
Computational improvements for data folding have been suggested~\cite{ain2018pystoch} as well.

\section{\label{sec:acknowledgements} Acknowledgements}
We appreciate the help of Sharan Banagiri, Patrick Meyers and Michael Coughlin during the process of data analysis software development.
We also thank Anirban Ain, Jishnu Suresh, Andrew Matas, and Sanjit Mitra for discussions regarding data folding.
The authors thank to the LIGO Scientific Collaboration for access to the data and gratefully acknowledge the support of the United States National Science Foundation (NSF) for the construction and operation of the LIGO Laboratory and Advanced LIGO as well as the Science and Technology Facilities Council (STFC) of the United Kingdom, and the Max-Planck-Society (MPS) for support of the construction of Advanced LIGO. Additional support for Advanced LIGO was provided by the Australian Research Council.
BG and ET are supported by ARC CE170100004.
ET is additionally supported by ARC  FT150100281.
This manuscript is LIGO-P1800105.
\bibliography{mybib}{}
\bibliographystyle{plain}

\newpage
\onecolumngrid
\section{\label{sec:appendix} Appendix}
\begin{figure}[!htb]
    \centering
    \begin{subfigure}[b]{0.49\textwidth}
        \includegraphics[width=\textwidth]{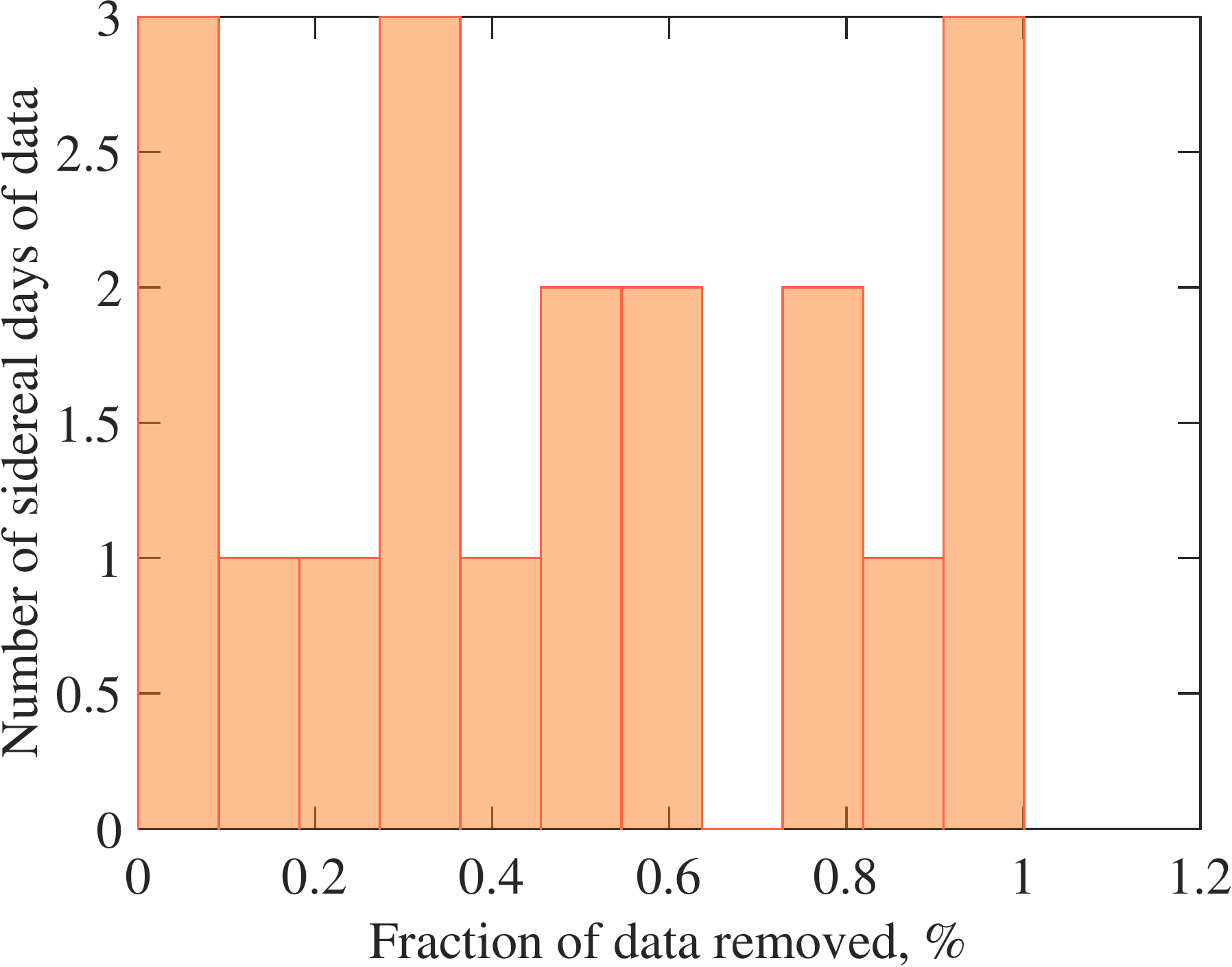}
       \caption{}
        \label{fig:daycutstats}
    \end{subfigure}
    ~ 
    \begin{subfigure}[b]{0.49\textwidth}
        \includegraphics[width=\textwidth]{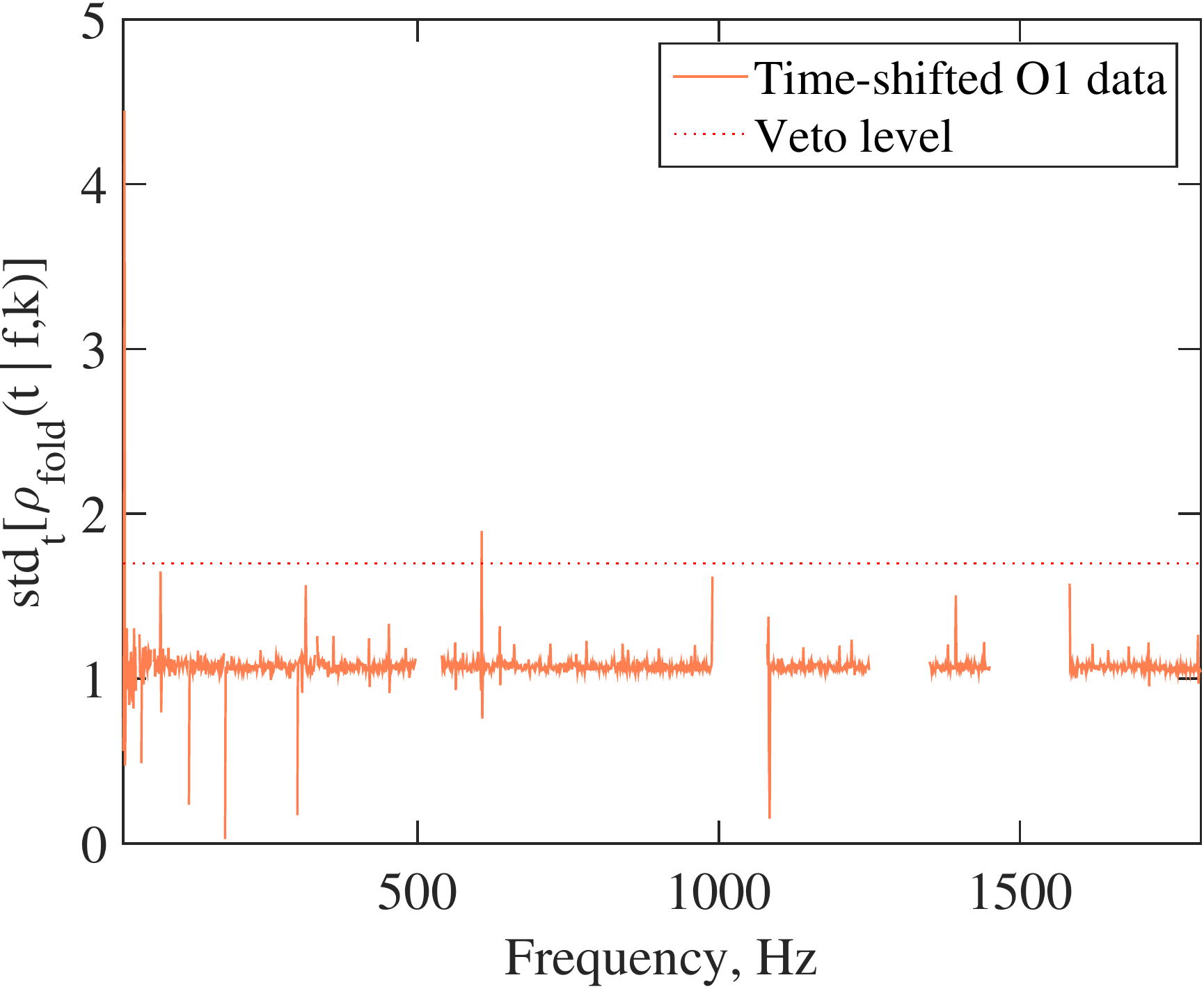}
        \caption{}
        \label{fig:stdrerho}
    \end{subfigure}
    ~ 
    \caption{Figure~\ref{fig:daycutstats} represents amounts of data removed from 14 days of time-shifted LIGO O1 data on the second data quality cut, described in Section~\ref{sec:dataquality}, that removes time segments. Figure~\ref{fig:stdrerho} represents the standard deviation of $\rho_{\text{fold}}(t|f,k)$ for time-shifted data from LIGO's O1 run.}
    \label{fig:additional}
\end{figure}
\end{document}